\documentclass[sigconf]{acmart}

\def\BibTeX{{\rm B\kern-.05em{\sc i\kern-.025em b}\kern-.08emT\kern-.1667em\lower.7ex\hbox{E}\kern-.125emX}}
    
\copyrightyear{2019}
\acmYear{2019}
\setcopyright{acmlicensed}
\acmConference[PEARC '19]{PEARC10 Conference}{July 28--August 01, 2019}{Chicago, IL}
\acmBooktitle{PEARC19 Conference, July 28--August 01, 2019, Chicago, IL}

\usepackage{verbatim}

\begin{document}

%
\title[Software-Enhanced Capabilities of an Ultra-High-Resolution Video Wall]{Software-Enhanced Teaching and Visualization Capabilities of an Ultra-High-Resolution Video Wall}
%
\author{Ramses van Zon}
\email{rzon@scinet.utoronto.ca}
\author{Marcelo Ponce}
\email{mponce@scinet.utoronto.ca}
\affiliation{%
  \institution{SciNet HPC Consortium, University of Toronto}
  \streetaddress{661 University Ave, suite 1140}
  \city{Toronto}
  \country{Canada}
  \state{Ontario}
  \postcode{M5G 1M1}
}

%
\renewcommand{\shortauthors}{Van Zon and Ponce}

%
\begin{abstract}
This paper presents a modular approach to enhance the
capabilities and features of a visualization and teaching room
using software. This approach was applied to a room with a large, high resolution (7680$\times$4320 pixels), tiled screen
  of 13 $\times$ 7.5 feet as its main display, and with a
    variety of audio and video inputs, connected over a network.
  Many of the techniques described are
  possible because of a software-enhanced setup, utilizing existing
  hardware and a collection of mostly open-source tools,
  allowing to perform collaborative, high-resolution visualizations as
  well as broadcasting and recording workshops and lectures. The
  software approach is flexible and allows one to add functionality
  without changing the hardware.
\end{abstract}

%
%


%

%
\maketitle

\section{Introduction}
In the summer of 2016, the offices of the SciNet HPC Consortium at the
University of Toronto in Canada were moved to a new location.  Its
previous location had a boardroom which also served as a video
conferencing room, as well as the room for training and courses, which
would usually be recorded and then posted, and sometimes also
broadcast.  But because of increasing demand for SciNet's training
workshops and courses, the room was almost never large enough. In
addition, there never was a proper space for visualizations, as the
screens in that room were neither particularly large nor of
particularly high resolution.

As we were designing our new office space, we had the vision of a
collaborative visualization space with a large, tiled,
ultra-high-resolution display.
The utilization of ultra high-definition video walls in different scientific fields offers
several advantages in many different aspects \cite{majumder2013large,2014PASA...31...33M},
ranging from higher resolution definition visualizations to sharpening wider fields of view.
The same room would also be used as a
teaching room, while a separate boardroom across the hall could mirror
the session and thus serve as an overflow room.  And because SciNet
serves a geographically distributed user base, it should be possible
for lectures, training sessions and meetings to be recorded and broadcast.

While it was clear from the onset that supporting
software would be needed to get all envisioned functionality, the
scope of the multi-server software setup turned out to be much
larger and involved than anticipated.

The hardware components provide a
100 square feet, 8K capable tiled display in a new
teaching/visualization room, which can switch between being driven
from a Windows PC server and a dedicated visualization GPU server running Linux.
Cameras were installed in the front and the back of the room, with
zoom and pan capabilities through a control screen in a lecture
console. The same console screen can mirror the large screen.  The lecture
console has microphones and there are wireless microphones for the room.
Sharing of the screens and of sound is possible with the boardroom's
two 4K monitors.  There are also inputs for laptops, Apple TV and
Chrome-cast.  In addition, there is a projector screen, and a
possibility for later hook up of a standard university teaching station PC, which
would allow this room to be used as a classroom by other
groups within the University of Toronto.

Not all of the envisioned capabilities
were realized out of the box. For instance, the maximum resolution of
the PC server or external laptops that could make it through the
various hardware layers and cables, is merely 1080 vertical lines.
The visualization server, meanwhile, can run in full 8K mode because its
output bypasses the hardware interface in charge of multiplexing
(i.e., choosing between the projector or video wall and sharing),
but that meant it could not be shown on
the preview window in the lectern console. The solution contained no
component that allowed online streaming of video and sound;
this would require another dedicated hardware appliance.

We used a variety of software to enable most of the envisioned
capabilities within the given hardware setup. This paper explains the
software stack utilized for driving the video wall, and for recording
and broadcasting. It ends by describing the several use cases of the current setup.
In our experience, vendors of A/V equipment are not
  necessarily aware of the software possibilities, and the modular,
  multi-server setup presented here appears to be novel.
We hope it can be of help to places that would like to set up a
similar facility.

\begin{figure}

\includegraphics[width=\columnwidth]{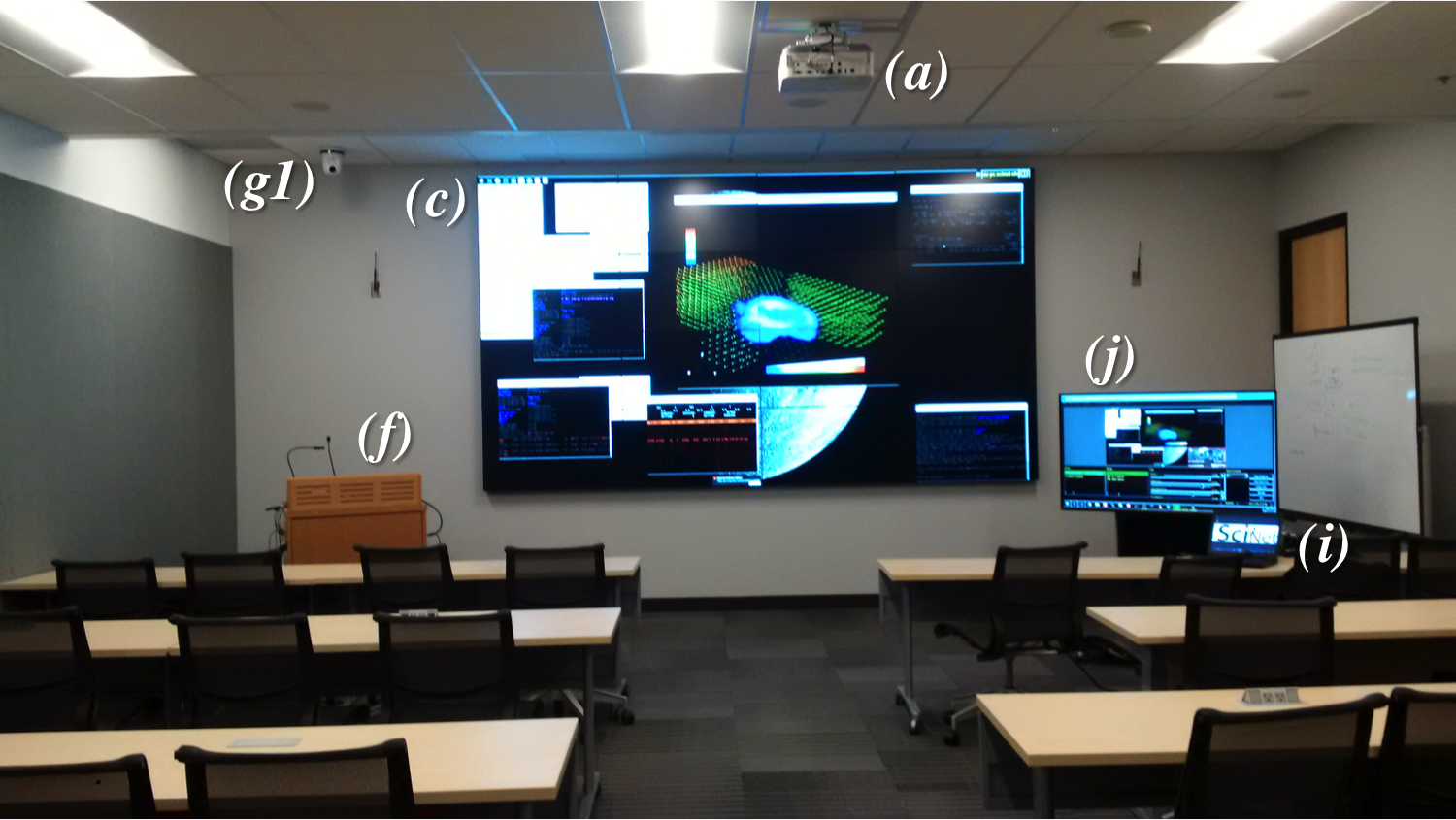}

\caption{A view of SciNet's teaching-visualization room located at SciNet's
headquarters in the University of Toronto St.\ George campus.}
 \label{fig:viz-room}
\end{figure}

\section{Hardware Components of the Visualization and Teaching Room}

To understand how the software running on the server(s) of the
  teaching room facilitates the room's
  functionalities, we will first describe the hardware setup.

\autoref{fig:viz-room} shows a picture of the teaching room.
A schematic diagram of the hardware layout of the visualization system
is shown in \autoref{fig:hardware-connectivity-diag}.

The most visible part of the hardware setup is the video wall
system ($c$). This is a Clarity Matrix LCD Video Wall 
\cite{ClarityMatrix1} by
Planar~\cite{Planar} composed of a
4 by 4 grid of 46 inch displays
with a small bezel
\cite{ClarityMatrix2}.

The total resolution is
7680$\times$4320 pixels (i.e., 8K or UHD resolution).
To supply content to the 16 tiles, they are grouped in units of 2$\times$2
tiles. Each unit acts as a single display; this is
accomplished by four hardware splitters
($b$).  As a result, the video wall can be driven as if it were a 2$\times$2
tiled display, and that allows the whole video wall to be driven from
just one high-performance graphics card with 4 output ports, hosted by
a single high-end visualization server, avoiding the need for a cluster to
drive the display (in contrast to e.g. Ref.~\cite{Johnson2012DisplayClusterAI}).
\footnote{The visualization server ($d$) contains two 14-core Intel Xeon CPUs
(E5-2680v4) at 2.40GHz, has 256GB of RAM, and an NVIDIA Quadro M400 graphics card, which can handle up to
four outputs.  The operating system of the visualization server is
\emph{openSUSE Leap}~42.2.}

In addition to the dedicated visualization server that directly drives
the video wall at its full resolution, it can also be driven by a
Crestron Matrix Switcher.
Because the tiled display
is also called a matrix, we will refer to the Crestron Matrix Switcher
simply as the \emph{Switcher}. The Switcher has multiple inputs
and outputs, which can be controlled from the console.

From the point of the Switcher, the video wall is
just one of the outputs. Other available output devices connected to the
Switcher are a projector ($a$) in the teaching room, and the displays in the
boardroom. The Switcher can work in a sharing mode as well, where the
audio and video from the teaching room are also played in the
boardroom, or vice versa.

The switcher system was designed to be HD capable in order to be compatible
with many input and output devices, but this does mean it currently
cannot redirect 8K or even 4K output to the video wall. This is the
reason the visualization server is directly connected to the video
wall instead of also being routed through the switcher, but that in
turn means the output of the visualization server cannot be shared
with the boardroom nor shown on the projector screen (devices which
are presently not 8K capable at any rate).

Other possible inputs to the Switcher are the cameras ($g1,g2$), laptops, Apple-TV and Google
Chrome-cast, as well as a PC server ($e$) intended for presentations and
applications that either requires \emph{Microsoft Windows} or needs to be shown on the
projector screen.\footnote{The PC server has two 4-core Intel Xeon CPUs (E5-2637v4) at 3.50GHz,
with 128GB of RAM, and runs \emph{Microsoft Windows} 10 as its operating system.}


\section{Required Functionality}

The need for an additional software layer arises from the different
kinds of usage that the system needs to support. Among the most common
usages are:

\begin{enumerate}

 \item High-resolution graphics: Using the visualization
   server to generate images or animations of data
   visualizations. This is mostly useful for research and outreach.

  \item Lectures and seminars: The video wall shows presentation
    material. Optionally, the lecture can be shared with the
    boardroom, or a second display could show presentation controls.
    Lectures are still a frequently used educational vehicle.

  \item Recording: like lecturing with added sound capture and
    optionally including the camera showing the lecturer.
    Requires post-processing to produce the video to be
    disseminated online.
    Such recordings are frequently requested by students,
      especially in lecture series, where they may have missed some
      lectures. Recordings can also form the basis of producing an
      online course or inverted classroom setup.
    
  \item On-line lecturing (or broadcasting): Instead of recording, the
    presentation and the camera output need to be combined and made
    available as an online stream in some standard format.
    Broadcasting is particularly useful when the
      audience of a lecture or seminar is geographically dispersed.
    
  \item Video conferencing: This requires a
    two-way communication of video and optional screen
    sharing. It is sometimes desirable to have multiple displays;
    one for the presentation material and one to see the presenter.
    Apart from video meetings, this capability is also useful
      in remote instruction.

  \item Visualization collaborations or demonstrations: One or several
    people might want to inspect and interact with an 8K visualization
    on the video wall. In case of a demonstration, recording should
    also be possible.
    This can be useful for research, training, and outreach.

  \item Handling multiple screen outputs: In addition to the large
    UHD video wall, in some cases it could be useful to count with
    additional screens to further increase the displaying real state.

\end{enumerate}

In the next section, we will explain how our audio-visual
  setup that was designed for only some of these functions, could be
  extended to provide all desired functionality.

\section{Modular Software Approach}
\label{sec:softsoln}

In order to fully exploit the features and
capabilities of the visualization and teaching room, we
used a series of software packages to augment the hardware
capabilities, which we will describe in this section.

The software solution relies heavily on a (wired)
network.  Through the network, video and control signals can be
exchanged between various servers (e.g.\ GPU server, PC, laptops,
and mobile devices), using software solutions such as
VNC\cite{vnc_IEEE_1998,x11vnc}, OBS\cite{OBS}, and
nginx-rtmp\cite{nginxrtmp}.
Wireless capabilities are added through a dedicated router 
(so that connected devices live on the same network and are not blocked
by firewall rules).

The first three ways of using the system described in the
  previous sections, i.e., high resolution graphics, lecturing and
  recording, require very few additional
software tools, as long as lecturing and recording at 1920$\times$1080 resolution
is acceptable. The camera feed is accessible from the PC server, on
which common presentation software as well as recording software can
be run. For recording, we at first installed
\emph{Camtasia}~\cite{Camtasia} (which has useful post-processing
features), but later also the free software
\emph{ActivePresenter}~\cite{ActivePresenter} and
the open-source \emph{Open Broadcaster Software}~\cite{OBS}.

The \emph{Open Broadcaster Software} also enables the fourth point,
online lecturing,
except for a final step: to broadcast to the internet, one needs
the feed from the software to be pointed to a \emph{streaming server}
which can then stream it through a standard protocol. Some streaming
servers are available online, but require at least an account. To have
our setup be independent of an external site, we instead installed an
\emph{nginx}-based media streaming server, the \emph{nginx-rtmp}
module (\emph{nginx} version 1.10, \emph{nginx-rtmp} version 1.13)~\cite{nginxrtmp}, inside a
\emph{Ubuntu} 16.04 virtual machine that runs on the PC server using
\emph{VirtualBox}~\cite{VirtualBox-website}.  This same module allows
streaming and recording at the same time.\footnote{The default
  streaming protocol of this module is ``rtmp'', which is associated with
  flash and not universally supported anymore, but it should be
  possible to switch to different protocols like ``hls'' and ``dash.''}

Video conferencing or remote participation, the fifth
  required functionality from the previous section, is possible with the
PC server through a variety of proprietary applications, such as
Skype, Vidyo and Zoom. This requires no further help from software, unless
one wants to have a two-monitor setup, one with slides and one with the
presenter, for instance 
(we will come back to this capability below).

The case of collaborative visualization or demonstrations
turned out to be the most involved to get to work.  In this case, a
group of people might be interactively involved in the visualization,
perhaps walking around and inspecting different parts of the displayed
graphics, or one person is interacting with the visualization while
explaining it to an audience. One might want to zoom in on particular
aspects, or change parameters, while looking at the video wall. The
challenge is to overcome the restriction that the visualization server
can only be controlled from the console, so one could not walk around
and manipulate the graphics.  A further complication is that the
console faces away from the video wall and doesn't show a preview of
the screen, because the console can only show output from the
Switcher, which the visualization server bypasses.

\begin{figure*}
\centering
\includegraphics[width=\textwidth]{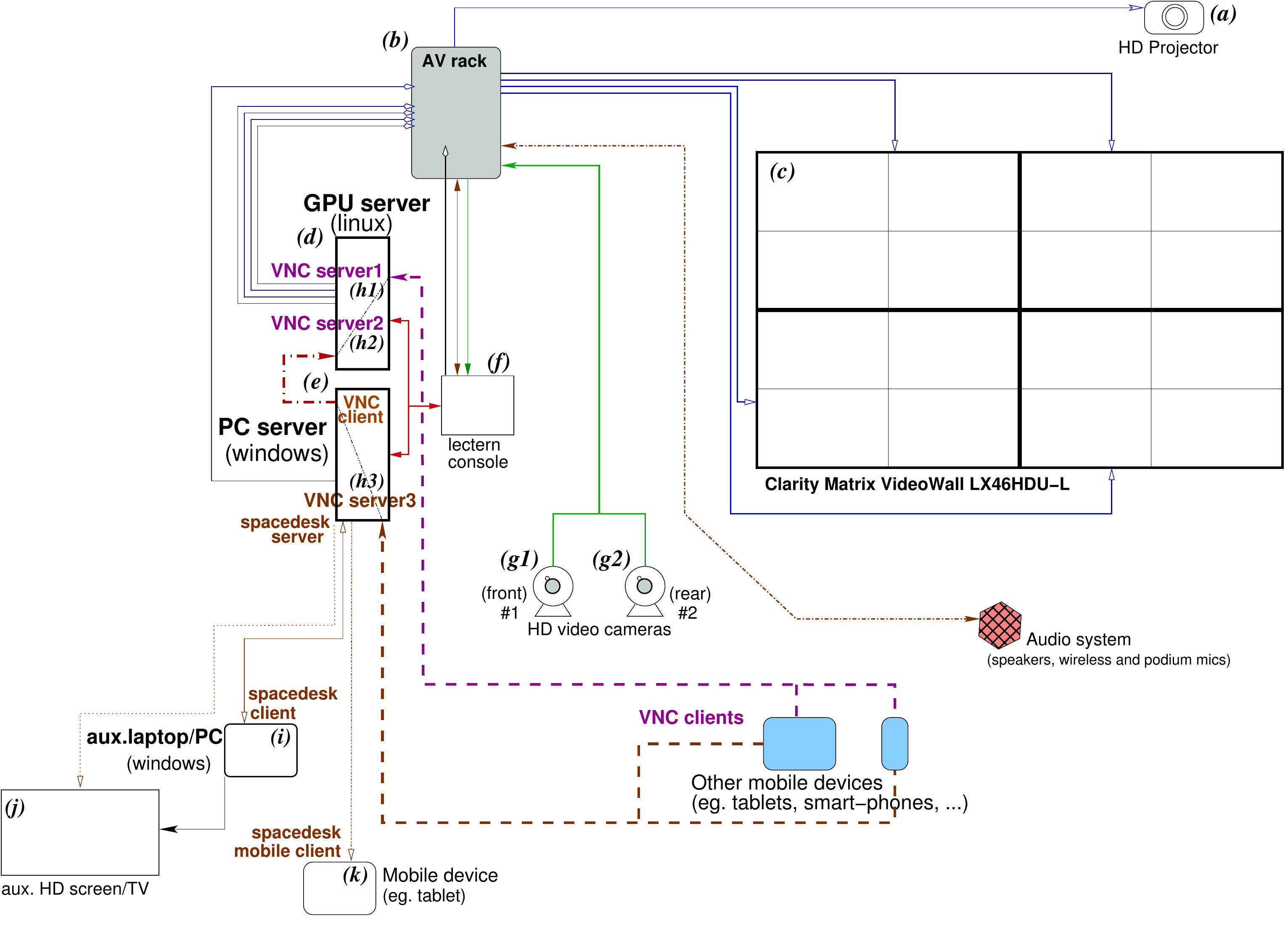}
\caption{Diagram of our video wall system, including
computer servers, audiovisual media, as well as software connection using VNC, denoted by dashed lines in the figure.
Notice that the AV rack, in addition to provide the hardware infrastructure
for connecting all possible medias, also serves as large multiplexer
allowing to share some of the different resources among the GPU and PC servers.
}
 \label{fig:hardware-connectivity-diag}
\end{figure*}

To make the setup suitable for such interactive usage, we first start
a Virtual Network Computing (VNC) server on the 8K visualization
server (``VNC server1'' in \autoref{fig:hardware-connectivity-diag}).  Because
VNC~\cite{vnc_IEEE_1998} is a protocol to view and interact with a remote screen,
where the VNC server captures the screen and sends this to VNC clients
on other computers, one can have a tablet
running the VNC client showing and controlling the same screen that is
showing on the video wall.

The VNC server on the visualization server that we picked for this was
\emph{x11vnc}~\cite{x11vnc}, which is capable of server-side scaling on Linux,
thus bringing the transfer rate needed to mirror the 8K desktop
image down to something that a laptop or tablet
can handle over a typical wireless network.\footnote{The
  PC server can be driven remotely as well, using a VNC server for
  \emph{Microsoft Windows}, such as \emph{UltraVNC} (``VNC server3'' in \autoref{fig:hardware-connectivity-diag}).}  The laptop or tablet can
then be used to control the video wall without being tethered to the
lecture console. A second scaling VNC server (``VNC server2'' in \autoref{fig:hardware-connectivity-diag})
on the GPU servers sends a 
 signal to a VNC client
on the windows PC that matches the PC's resolution. The Windows PC also captures the camera input, and
runs the capturing software, enabling a recording of the event
at the lower resolution of 1920$\times$1080,
while the video wall showed the full 8K visualization.
In addition, the PC server's screen was visible on the lecture
console, so lecture-style demonstrations were still possible.
See \autoref{fig:hardware-connectivity-diag} for a detailed illustration.

Finally, the addition of further
  displays or monitors can be of use in presentations and remote
    instruction. The hardware was not designed with this in mind, but
    it is possible to do this in software, such that the system could
    have a secondary monitor which is, e.g., 
a 55" screen or a smaller computer monitor (see $(j)$ in \autoref{fig:viz-room}).
The first option (an additional monitor) can be useful in the case of having a remote participation
or demonstration, where the video wall is used to display the main presentation
or participation, whereas the additional screen can be used to show the presenter
or other details related to the main visualization.
The second setup (using an additional laptop) can be useful as an aid to the presenter, where information about the
active presentation in the video wall is visualized
privately on the laptop for the presenter,
eg. slideshows in presenter mode, or confidential material not to be shared through the videowall.

In order to achieve this, a Windows laptop was added to the
  setup $(i)$, which was connected to the additional display $(j)$. We use a
  software named spacedesk\cite{spacedesk}. The spacedesk server
  portion runs on the Windows server $(e)$, while the spacedesk client runs
  on the laptop $(i)$.  What spacedesk accomplishes is that the client's
  display becomes visible as a second monitor on the Windows
  server. To be able to run spacedesk, there are restrictions on the
  server side: The spacedesk server requires Windows 10 as its
  operating system (which the Windows server has).  However, the
  client (here, the additional laptop) can run an older version of Windows, an Apple device, or an
  android device. The second display could therefore even be a portal
  mobile device, such a tablet or a smartphone $(k)$.

\begin{table*}
\centering
\begin{tabular}{| l || l | l |}
	\hline
		server		&	software stack		&	accessible hardware 		\\
	\hline\hline
	{\it PC server (e)}	&	VNC server		&	videowall $(c)$ (with 2K full-resolution)	\\
	OS: Windows		&	VNC client		&	HD cameras $(g1,g2)$	\\
				&	VirtualBox:
						Linux server/streaming
								&	audio system	\\
				&	OBS			&	HD projector $(a)$	\\
				&	Skype/Vidyo		&	USB inputs (wireless mic, wireless keyboard and mouse, ...)	\\
				&	spaceDesk server	&	\\
	\hline
	{\it GPU server (d)}	& VNC server: x11vnc		&	videowall $(c)$ (with 8K full-resolution)	\\
	OS: Linux		& Visualization packages:
					VisIt, ParaView, VMD, Gephi, ...	
								&	audio system		\\
				& Video manipulation packages:
					ffmpeg, mencoder, VLC, ...
								&	USB inputs		\\
				& NVIDIA drivers, openGL, ...	&	\\
	\hline\hline
\end{tabular}
\caption{Summary of servers, most relevant components from the software stack and corresponding accessible hardware capabilities.}
\label{table:servers-software}
\end{table*}


The result of combining, VNC, OBS, nginx-rtmp, and other
  software, is that all six capabilities from the previous section are
  realized in the teaching/visualization room.
\autoref{table:servers-software} recaps each element in the software stack in correspondance with the servers and its accesible hardware functionalities.
\\
There are other software approaches/systems to drive and utilize large video walls.
All of them present substantially different scopes, goals and higher
complexity to the solution presented here.
For instance,
  \emph{SAGE2} --Scalable Amplified Group Environment-- \cite{marrinan2014sage2}
  from the Electronic Visualization Laboratory at the University of Illinois at Chicago (EVL/UIC),
is an interesting tool for sharing files and images through a web-browser interface.
Its utilization is quite simple from the user perspective, however
its main and default features are mostly limited to upload and manipulate certain
predefined file types (images, videos, documents) into the video wall.
It is also possible to write specific applications for taking advantage of \emph{SAGE2},
but complex software and complicated software interfaces/interactions,
such as the ones described in the following section would not be possible to achieve in this framework.
\emph{Display Cluster} \cite{Johnson2012DisplayClusterAI} from Texas Advanced Computing Center (TACC),
offers a more complex and elaborated approach to driving a tiled display, fulfilling these and other features.
However it requires a hardware setup much more sophisticated than the one presented here,
basically it is driven by a cluster-type server.
Moreover, we should emphasize that the type of hardware, ie. video wall we have,
has to be handled as just one input/output connection due to
 a design requirement to be able to connect many
  different input devices to the system.


\section{Demonstrations}
\label{sec:demos}
We will briefly describe a few cases to demonstrate the capabilities of the video wall utilizing the
software solution described in \autoref{sec:softsoln}.
A full recording of 
these demonstrations
is available on our YouTube channel \cite{SciNetYouTubeChannel}. 
These were part of the grand opening of the new office space.
Some snapshots of these demonstrations, taken utilizing the setup described
in the previous section, are displayed in 
\autoref{fig:gephi-demo} and \autoref{fig:sph-NSBH-demo}.
Two demos
showed how to get insights into complex data: in the first case, by visualizing the software installed
on SciNet's main cluster by using graph visualizations (\autoref{fig:gephi-demo}); 
in the second one, by presenting several simulations generated using computational astrophysics,
e.g. interacting neutron stars, with data from numerical general relativity simulations.

A third demonstration of the room's capabilities are in usage
as a host site in a remote instruction event.
%
We should also note that the teaching/visualization room is actively
used in the education and training program of
SciNet\cite{SciNetWhitepaper}. While most events in this program are
face-to-face, lectures are broadcast, recorded and posted on-line as
well on our education website \cite{SciNetEdu}.


\subsection{Graph visualization for software dependencies}
\label{sec:gephi-demo}
The video wall can be used to explore and probe large data sets that
can be represented as graphs.  As an example, we considered the full
corpus of software tools available on SciNet's General Purpose Cluster (GPC)
\cite{1742-6596-256-1-012026}.  The GPC has been operating for eight 
years with a large user base with diverging and conflicting software
needs. These needs are met by providing a bare-bones Linux operating
system supplemented by \emph{environment modules}~\cite{modules1,modules2,modules3} 
that can be loaded on demand by users to provide the particular software
they need. In the module system, conflicts can be defined between
modules to ensure that incompatible software is not loaded at the same
time, while requirements can be defined to make sure software
dependencies are met. Several versions of software packages are often
available. Old modules are not removed (for reproducibility), causing
the module system to have grown into a sizable complex network.

In this demonstration, we scraped the whole set of software modules on the GPC and represented the data as a
network where each node is a module and the lines of different colours
represent either dependencies between different packages, conflicts
among them, or likeness (e.g. versions of the same software).

\begin{figure}
\centering
  \includegraphics[width=\columnwidth]{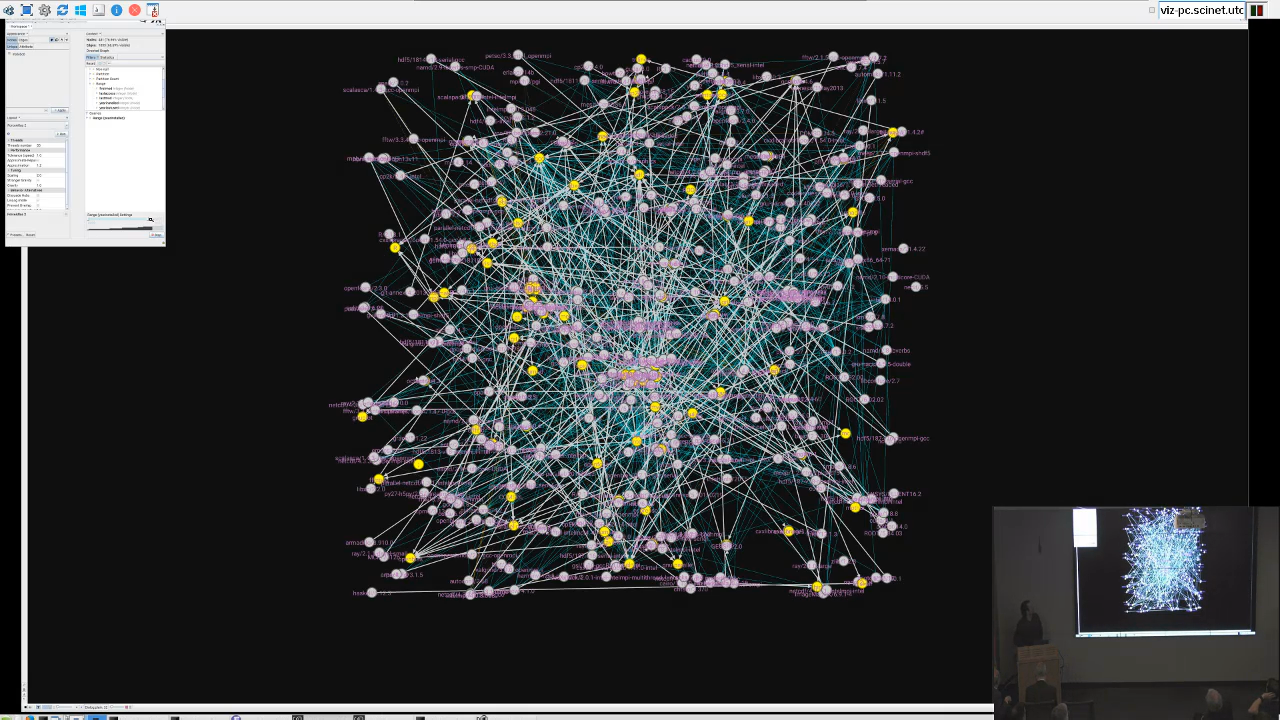}
\caption{
Snapshot of a live demonstration using the videowall and inset with the presenter,
showing how the different presentation sources can be combined and captured for
recording and/or broadcasting purposes.
The central image represents an
 8K UHD snapshot of the software module system of
  the SciNet GPC cluster~\cite{1742-6596-256-1-012026} as visualized by
  \emph{Gephi}~\cite{ICWSM09154}.  Each node represents a particular software
  module on the GPC cluster, cyan lines represent dependencies while
  orange ones represent conflicts, meaning software modules that
  cannot be loaded at the same time.  The graph representation tends
  to aggregate similar modules together based on edge weights, using \emph{Gephi}'s ``ForceAtlas
  2'' layout algorithm \cite{jacomy14}.  The
  inset is a zoom-in of a region. This is included only for the
  purpose of showing details in this paper; when this graph is displayed on the
  video wall, one does not need to zoom in, one can simply move closer
  to the screen to see those details.}
 \label{fig:gephi-demo}
\end{figure}

The visualization is done using \emph{Gephi} \cite{ICWSM09154} (version 0.9.1), which 
allows to interact in real time with the network representation.
This is one of the best cases and examples for utilizing this type of
video wall as a display, as it allows the user to have a great 'big
picture' view of the data being displayed (see
\autoref{fig:gephi-demo}) while also focusing on the details which are
sharpened by the high resolution of the display (see inset in
\autoref{fig:gephi-demo}). The visualization is dynamic: one can
change the node-placement algorithm, one can select subsets of nodes
based on attributes (e.g., to only show software modules available in a
specific past year), hide and show labels, etc.

The node placement used here is
  \emph{ForceAtlas2}\cite{jacomy14}, a force-based algorithm. Gephi
  uses OpenGL to render its graphics.

\subsection{Visualization of different 
  astrophysical simulations}
\begin{figure}
\centering
  \includegraphics[width=\columnwidth]{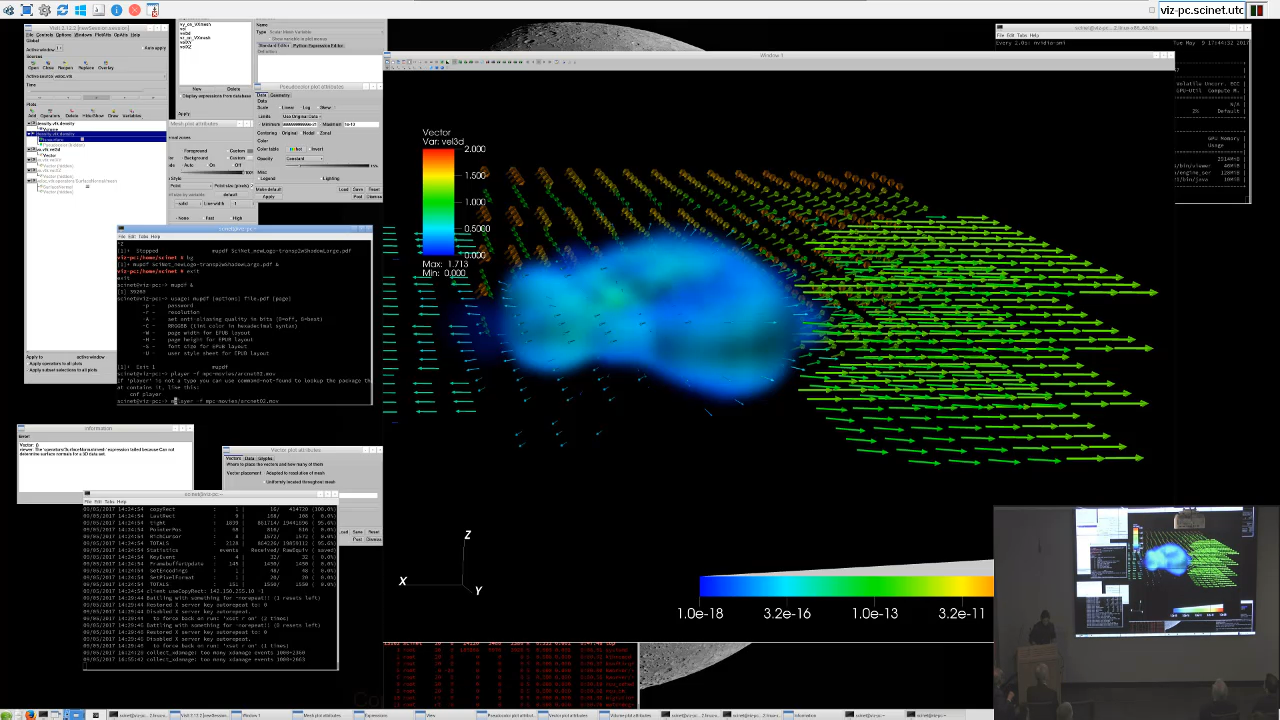}
\caption{
Snapshot of another live demonstration using the videowall and inset with the presenter,
showing how the different presentation sources can be combined and captured for
recording and/or broadcasting purposes.
The main screen displays
 an 8K UHD image generated using
 \emph{VisIt}~\cite{HPV:VisIt}. 
The image combine different visualization techniques such as isosurfaces,
transparencies, streamlines, and velocities fields.
}
 \label{fig:sph-NSBH-demo}
\end{figure}

The astrophysical visualizations demonstrations were generated as high-resolution
videos presenting results from different study cases: gravitational recoil of
black holes surrounded by accretion disks \cite{Ponce:2011kv}, interaction of
magnetized neutron star binaries \cite{Palenzuela:2013hu,Ponce:2014sza}, and
tidal disruption events of neutron stars by black holes \cite{Roberts:2016igt}.
The videos were generated utilizing the following open source visualization
tools: \emph{SPLASH} \cite{price_2007} for the Smoothed-Particle Hydrodynamics (SPH)
simulations used for modelling the effects on the accretion disk and tidal
disrupted events, and \emph{VisIt} \cite{HPV:VisIt} for the binary neutron star
interactions visualizing the 3D magnetic field lines interactions, as well as,
current sheets and star fluid density.
\autoref{fig:sph-NSBH-demo} shows the result of such a visualization
from the data of a simulation of a neutron star disrupted
by the interaction with a black hole, where several
numerical relativity techniques were utilized \cite{Roberts:2016igt}.
Moreover this particular visualization combined an interpolation of the SPH particles
and velocity fields into a 3D Cartesian grid that were imported into \emph{VisIt}, to generate
representations of density profiles, streamlines, and velocity fields.
%
The interpolated fields were saved using the VTK legacy format\footnote{\url{http://www.vtk.org/wp-content/uploads/2015/04/file-formats.pdf}},
so that these can be imported into \emph{VisIt}, where the respective physical
quantities are represented via pseudo-color plots, iso-surfaces and iso-volume slices for the
fluid density and glyphs and streamlines for the velocity field.
%
%
As it is usually the case, the visualization of streamlines and vector field representations are in big part
an art and strongly dependent on the geometry of the problem, as well as, the selection
of the location for the \textit{seeds} utilized for aiding the stream-or-field lines origins.
The streamlines representations were generated integrating in both directions in the pathline,
utilizing a \textit{Dormand-Prince} (Runge-Kutta) integrator, where tolerances and
maximum steps of integration were adjusted for each particular case.
In the case of the SPH simulation, the situation is easier as the selection is
guided by the interpolation into the Cartesian grid.
However, the magnetic field lines are in general way more challenging, not only for the
strong dynamical regime of the cases presented here, but also due to the intrinsic nature of
the magnetic field lines.
%

As \emph{VisIt} (as well as \emph{Gephi}) has the capability of generating images of
arbitrary resolution, generating images in full 8K 
resolution can be performed on any reasonably powerful
visualization server, but the video wall allows to view it at a scale
on which all details can be seen.


The final presentation for the astrophysical visualizations
was generated combining the individual videos of the
previous mentioned cases, sliding through them using \emph{Prezi}~\cite{prezi.com}
via their academic/educational license\footnote{A complete demo can be found in the following link:
\url{http://prezi.com/qv_ke57mkjri/?utm_campaign=share&utm_medium=copy}.}.
An alternative open-source tool with similar functionalities to those
of \emph{Prezi} is \emph{impress.js}\footnote{
\emph{impress.js} is a presentation framework based on CSS3 (see \url{https://github.com/impress/impress.js/})}
 which can be coded using simple html-type commands,
or, for instance, utilizing \emph{hovercraft} via its python wrapper interface.
%
%
%
%
%
%
%
\subsection{Collaborative workshop}
\label{sec:collab-workshop}
The preceding demonstrations were the very first actual uses
of our setup employing the software enhanced solution presented here.
Since then, we have also utilized this setup for broadcasting and
remotely hosting several workshops type of activities, e.g. \cite{PetaScaleInstitute,BrainHack,VizHack}.
For this type of events, we used the multi-monitor setup described in
\autoref{sec:softsoln} and shown in \autoref{fig:viz-room}, in
combination with Zoom\cite{zoom}. 
%
%
Without the advanced teaching/visualization room, these events could not
have been offered.


\section{Conclusion}
\label{sec:concl}

In this paper, we have presented a versatile and modular setup to
drive, manage and enhance the utilization of a video wall system in
its native UHD resolution for visualization, lecturing, broadcasting,
recording, video conferences, collaborations, and demonstrations.
We further described a number of demonstrations employing this setup.

The novelty we would like to emphasize here
is the relative simplicity, flexibility and cost efficient approach we proposed.
One limitation with the current setup is related to the precise GPU model
our server uses, which does not have the capability
of rendering 8K videos in hardware. But given the modularity and flexibility of
the current approach,
such a restriction can easily be overcome by upgrading the GPU model
that does have this capability.
Flexibility of the setup is also important to be able to
  incorporate new features and modern technology-enhanced teaching techniques.

This software-enhanced solution will be applicable to other systems
like this, if not in its entirety then at least in parts.  The
approach augmenting the existing capability of a video wall using
software, is more efficient in terms of cost and flexibility than
redesigning and upgrading the hardware system.

%
\begin{acks}
We want to thank all our colleagues at SciNet, especially
Daniel Gruner, Scott Northrup, Marco Saldarriaga and Jason Chong, whom helped
dealing with the logistics and setting up the hardware and servers in our
visualization-teaching room. We thank Leslie Groer and Erik Spence for
careful reading of the manuscript.
\end{acks}

%
\bibliographystyle{plain}
\bibliography{videowall-refs}

\end{document}